# Emotion-based Recommender System

Hao Wang
CEO Office
Ratidar Technologies LLC
*Beijing*, China
haow85@live.com

***Abstract*—Recommender system is one of the most critical technologies for large internet companies such as Amazon and TikTok. Although millions of users use recommender systems globally everyday, and indeed, much data analysis work has been done to improve the technical accuracy of the system, to our limited knowledge, there has been little attention paid to analysis of users' emotion in recommender systems. In this paper, we create a new theory and metrics that could capture users' emotion when they are interacting with recommender systems. We also provide effective and efficient visualization techniques for visualization of users' emotion and its change in the customers' lifetime cycle. In the end, we design a framework for emotion-based recommendation algorithms, illustrated in a straightforward example with experimental results to demonstrate the effectiveness of our new theory.**

## I. INTRODUCTION

Recommender system is an effective information retrieval tool that is capable of capturing users' preferences and predict future purchase. Compared with conventional or digital marketing campaigns, recommender system could increase website / app traffic by a large margin while the maintenance cost is only a small fraction of its alternatives. Since its debut in 1992, recommender system has served large internet companies such as Amazon and TikTok. The cost-effectiveness of the technology is paid huge attention to and the technology has received a tremendous amount of investment during the past decades.

One subfield of recommender system is context aware recommender system (CARS). For example, a built-in car music audio system might recommend music to passengers based on the scenary or their emotions. This type of emotion-based CARS system is a revolutionary idea, but due to the difficulty of acquiring human emotion data, there has been little success in materializing the idea of emotion-based CARS system. Smart computation of human emotion (emphatic computing) is critical in computer systems like the one just mentioned. Since it is so difficult to monitor human emotion in practice, in this paper we develop a new theory with new metrics without using sensors or other hardwares for emphatic computing with recommender systems.

The novelty of our methodology is that we do not rely on sensor data to do emphatic computing. All we need is users' online behavior - no more and no less than the data requirement for an ordinary recommender system. After presenting our methodology, we illustrate how to visualize users' emotion when using recommender system. In the end, we show how to build emotion-aware recommender systems with a new framework. We give an example, i.e., emotion-based matrix factorization to illustrate our idea. To the best of our limited knowledge, we are the first to conduct research of this type.

## II. RELATED WORK

Recommender system was first proposed by D. Goldberg in 1992 [1]. The algorithm was named collaborative filtering. Although a successful algorithm dominating the industry during the first years of recommender system history, in recent years, H. Wang [2] first analyzed the Matthew Effect and sparsity problem of the algorithm, and then proved that collaborative filtering is theoretically wrong [3][4].

Around 2007, with the popularity of the Netflix Movie Recommendation Contest, a new milestone, namely probabilistic matrix factorization [5] was introduced. The algorithm decomposes a full user item rating matrix into dot product of user feature vectors, and item feature vectors, just as its follow-up variants [6][7][8]. In 2011, S. Rendel [9] proposed a new algorithm named Factorization Machines, that could incorporate contextual information for recommendation. A variant of factorization machines named FFM [10] was later proposed and widely adopted in the industry.

Shortly after the debut of matrix factorization algorithms, learning to rank algorithms such as Bayesian Personalized Ranking [11] was invented to remodel the recommendation problem as a ranking problem rather than prediction problem. Other inventions in this school include Collaborative Less is More Filtering [12], Pareto Pairwise Ranking [13], Skellam Rank [14], etc. Research in this direction is continued to even today.

In the deep learning era, Wide & Deep [15], DLRM [16], and DeepFM [17], among a whole spectrum of algorithms, have become the de facto standard of industrial recommender systems. Research focus has diverged from technical accuracy in recent years to other problems such as fairness [18][19][20] and cold-start problem [21][22][23].

Emphatic computing is a relatively obscure research topic. Although researchers have conducted research on LLM [24][25], there has been little research focus on emphatic recommender system.

Identify applicable funding agency here. If none, delete this text box.

## III. EMPHATIC RECOMMENDER SYSTEM

When people talk about being emotional, we believe they are referring to the following phenomena :

- For a very **popular** item, if the user **dislikes** it, it means the user is emotionally biased **against** the item.
- For a very **obscure** item, if the user **likes** it, it means the user is emotionally biased **towards** the item.

By popular, we mean the item receives very high score by the mass overall, or the item receives lots of publicity no matter what its score is. By obscure, we mean the item receives very low score by the mass overall, or the item receives very little publicity no matter what its score is.

Based on the above observation, we define the Emotional Score (ES) of a recommender system as follows :

$$ES_{i,j} = \frac{1/R_{i,j}}{Score_j \times Count_j} I(Score_j \text{ is large or } Count_j \text{ is large}) + \frac{R_{i,j}}{Score_j \times Count_j} I(Score_j \text{ is small or } Count_j \text{ is small})$$

, where $ES_{i,j}$ is user i's emotion of item j, $R_{i,j}$ is user i's rating of item j, $Score_j$ is the average rating of item j, $Count_j$ is the number of item occurrence in the user item rating dataset, and I() is the identity function.

In implementation, in order to deal with the skewed data distribution, we use normalized $\log(ES_{i,j})$ instead of $ES_{i,j}$ itself to compute the emotional aspect of the user data.

## IV. DATA VISUALIZATION

We choose MovieLens 1 Million Dataset [26] and LDOS-CoMoDa Dataset [27] for our visualization experiment. We create the data matrix by computation of the Emotional Score of each user item pair in the dataset. When data is missing, we use 0 to replace the missing value. When use colormap to visualize the Emotional Scores of the datasets :

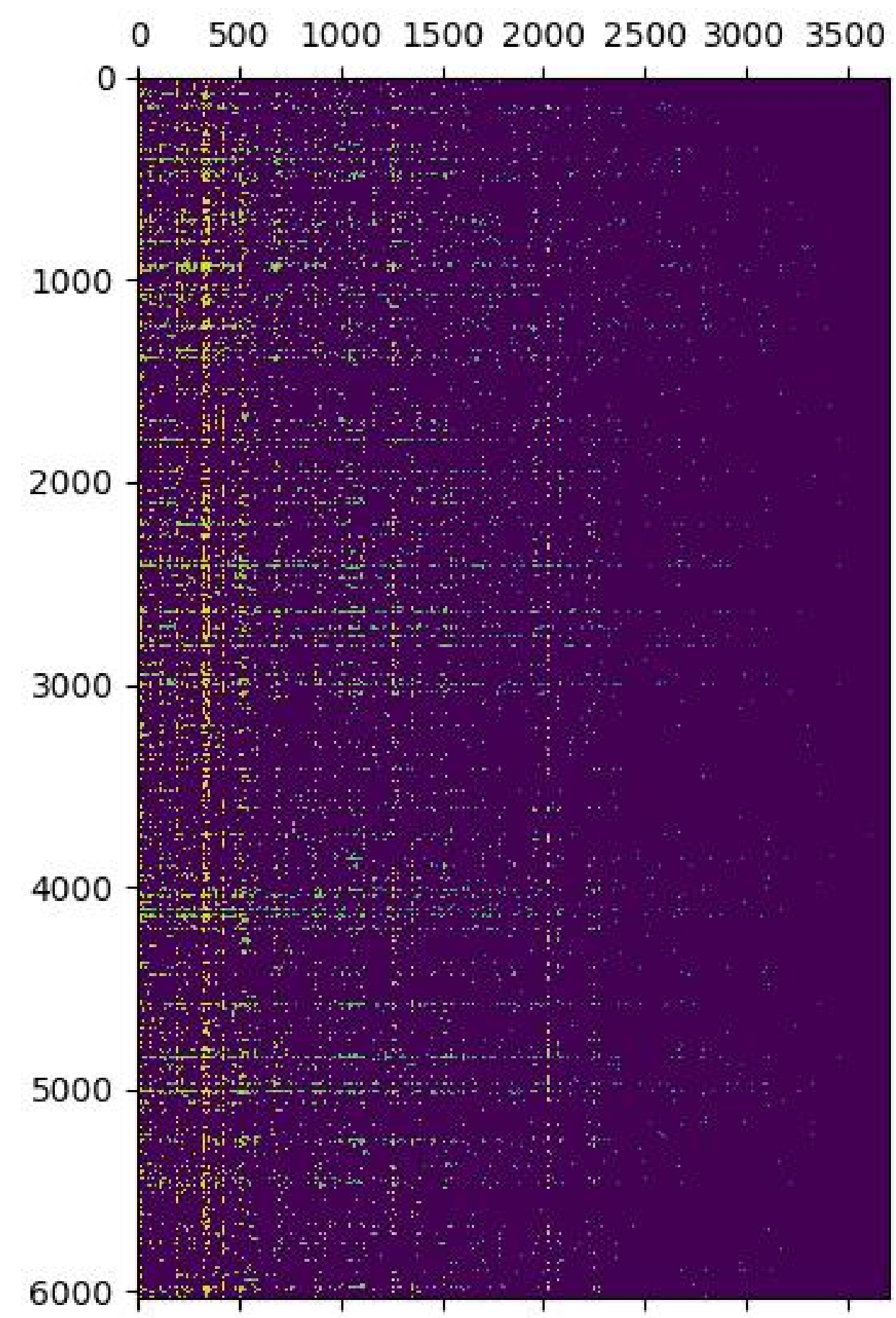

Figure 1 Emotional Scores of Users in MovieLens 1 Million Dataset. Rows represent users and columns represent items (movies).

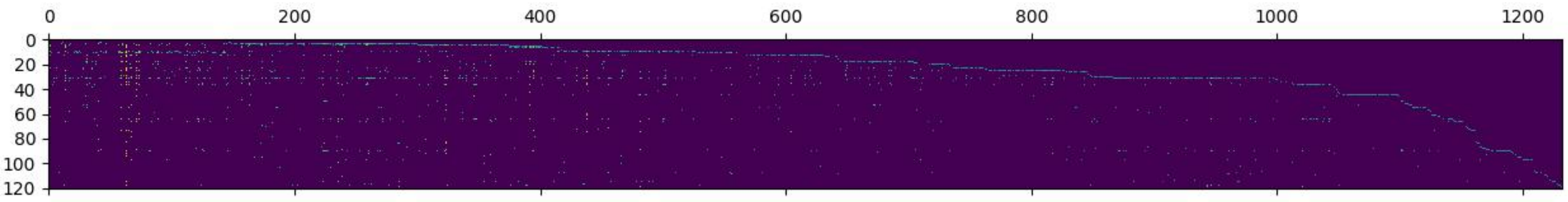

Figure 2 Emotional Score of Users in LDOS-CoMoDa Dataset. Rows represent users and columns represent items (movies).

Figure 1 clearly demonstrates the Matthew Effect of the Emotional Scores while Figure 2 shows a clear demarcation line (greenish blue) extending from the upper left corner to the lower right corner.

We computed the most emotional movies by computing the average Emotional Scores related to each movie, and obtained the following result (Top 10 most emotional movies):

| Title | Year | Genre |
| --- | --- | --- |
| American Beauty | 1999 | Comedy\|Drama |
| Star Wars: Episode IV - A New Hope | 1977 | Action\|Adventure\|Fantasy\|Sci-Fi |
| Star Wars: Episode V - The Empire Strikes Back | 1980 | Action\|Adventure\|Drama\|Sci-Fi\|War |
| Raiders of the Lost Ark | 1981 | Action\|Adventure |
| Saving Private Ryan | 1998 | Action\|Drama\|War |
| Silence of the Lambs, The | 1991 | Drama\|Thriller |
| Sixth Sense, The | 1999 | Thriller |
| Matrix, The | 1999 | Action\|Sci-Fi\|Thriller |
| Schindler's List | 1993 | Drama\|War |
| Shawshank Redemption, The | 1994 | Drama |

Table 1 Top 10 Most Emotional Movies
in the MovieLens 1 Million Dataset

From Table 1 we observed that the 5 out of the top 10 most emotional movies are action movies, which aligns with our intuition that action movies are more thrilling. The analysis of the LDOS-CoMoDa dataset is analogous, and we omit the details in this paper.

## V. EMOTION-BASED RECOMMENDATION

Let f be the loss function to be minimized for a recommendation algorithm, we define the loss function of emotion-based recommendation algorithm as follows:

$$L = f(R_{i,j}, \theta) - \lambda \sum_{i=1}^{N} \sum_{j=1}^{M} ES_{i,j}$$

, where $R_{i,j}$ is the user item rating value, $\theta$ is the parameters for the recommendation algorithm, and $\lambda$ is the constant regularization coefficient.

Let's take the example of matrix factorization and examine the technical performance of emotion-based recommendation. The loss function in this scenario is below:

$$L = \sum_{i=1}^{N} \sum_{j=1}^{M} \left( \frac{R_{i,j}}{\max(R_{i,j})} - \frac{U_i^T \cdot V_j}{||U_i^T|| \cdot ||V_j||} \right)^2 - \lambda \sum_{i=1}^{N} \sum_{j=1}^{M} ES_{i,j}$$

We further expand the loss function as follows :

$$L_{i,j} = \begin{cases} \left( \frac{R_{i,j}}{\max(R_{i,j})} - \frac{U_i^T \cdot V_j}{||U_i^T|| \cdot ||V_j||} \right)^2 - \lambda \frac{\frac{||U_i|| \times ||V_j||}{U_i^T \cdot V_j}}{Score_j \times Count_j} & , Score_j \text{ is large or } Count_j \text{ is large} \\ \left( \frac{R_{i,j}}{\max(R_{i,j})} - \frac{U_i^T \cdot V_j}{||U_i^T|| \cdot ||V_j||} \right)^2 - \lambda \frac{\frac{U_i^T \cdot V_j}{||U_i|| \times ||V_j|}}{Score_j \times Count_j} & , otherwise \end{cases}$$

We apply Stochastic Gradient Descent (SGD) algorithm to the emotion-based matrix factorization algorithm, and obtain the following formulas :

- When $Score_j$ is large or $Count_j$ is large :

$$U_i = U_i + \beta\left( \frac{2 \cdot \left( \frac{R_{i,j}}{max(R_{i,j})} - \frac{t_3}{t_2} \right)}{t_2} \cdot V_j - \frac{2 \cdot t_3 \cdot \left( \frac{R_{i,j}}{max(R_{i,j})} - \frac{t_3}{t_2} \right)}{t_0^3 \cdot t_1} \cdot U_i + \frac{B \cdot t_1}{t_3 \cdot t_0} \cdot U_i - \frac{B \cdot t_2}{t_3^2} \cdot V_j \right)$$

, where :

$$t_0 = ||U_i||, t_1 = ||V_j||, t_2 = t_0 \cdot t_1, t_3 = U_i^T \cdot V_j, B = \frac{\lambda}{Score_j \times Count_j}$$

$$V_j = V_j + \beta\left( \frac{t_4}{t_2} \cdot U_i - \frac{t_3 \cdot t_4}{t_0 \cdot t_1^3} \cdot V_j + \frac{B \cdot t_0}{t_3 \cdot t_1} \cdot V_j - \frac{B \cdot t_2}{t_3^2} \cdot U_i \right)$$

, where :

$$t_0 = ||U_i||, t_1 = ||V_j||, t_2 = t_0 \cdot t_1, t_3 = U_i^T \cdot V_j, t_4 = 2 \cdot \left( \frac{R_{i,j}}{max(R_{i,j})} - \frac{t_3}{t_2} \right), B = \frac{\lambda}{Score_j \times Count_j}$$

- In other scenarios :

$$U_i = U_i + \beta\left( \frac{2 \cdot \left( \frac{R_{i,j}}{max(R_{i,j})} - \frac{U_i^T \cdot V_j}{t_2} \right)}{t_2} \cdot V_j - \frac{2 \cdot t_3 \cdot \left( \frac{R_{i,j}}{max(R_{i,j})} - \frac{t_3}{t_2} \right)}{t_5 \cdot t_1} \cdot U_i + \frac{t_4}{t_0} \cdot V_j - \frac{t_3 \cdot t_4}{t_5} \cdot U_i \right)$$

, where :

$$t_0 = ||U_i||, t_1 = ||V_j||, t_2 = t_0 \cdot t_1, t_3 = U_i^T \cdot V_j, t_4 = \frac{\lambda}{Score_j \times Count_j} \cdot t_1, t_5 = t_0^3$$

$$V_j = V_j + \beta\left( \frac{t_4}{t_2} \cdot U_i - \frac{t_3 \cdot t_4}{t_0 \cdot t_1^3} \cdot V_j + \frac{B \cdot t_3}{t_2} \cdot V_j + \frac{B \cdot t_1}{t_0} \cdot U_i \right)$$

, where:

$$t_0 = ||U_i||, t_1 = ||V_j||, t_2 = t_0 \cdot t_1, t_3 = U_i^T \cdot V_j, t_4 = 2 \cdot \left( \frac{R_{i,j}}{max(R_{i,j})} - \frac{t_3}{t_2} \right), B = \frac{\lambda}{Score_j \times Count_j}$$

## VI. EXPERIMENT

We name our emotion-based recommender system algorithm EMF (Emotion-based Matrix Factorization), and compare its accuracy and fairness performance (Degree of Matthew Effect [28]) with ZeroMat [21], Random Placement, Classic Matrix Factorization [5], DotMat [29] and DotMat Hybrid [29] on MovieLens 1 Million Dataset [26] and LDOS-CoMoDa Dataset [27]. We obtained the following results (Figure 3 - Figure 5):

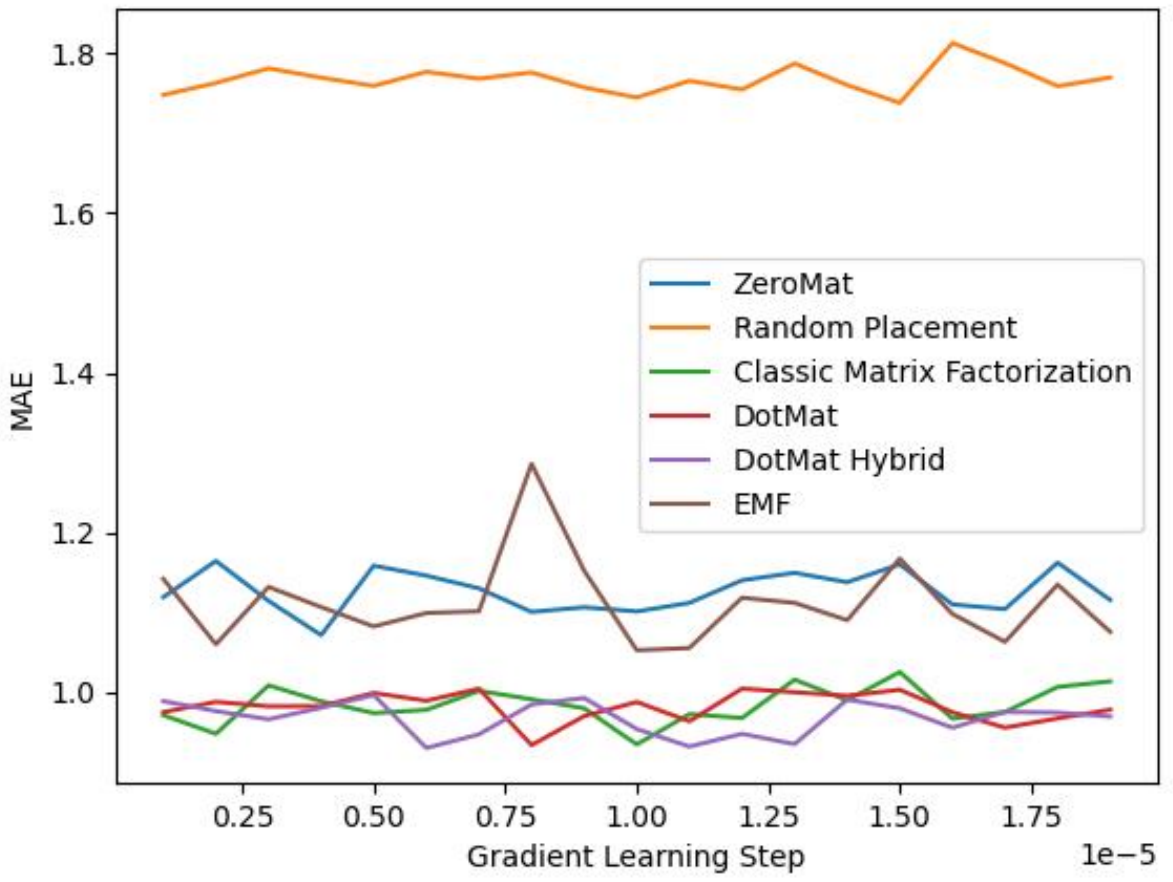

Figure 3 Comparison of algorithms on MovieLens 1 Million Dataset (MAE)

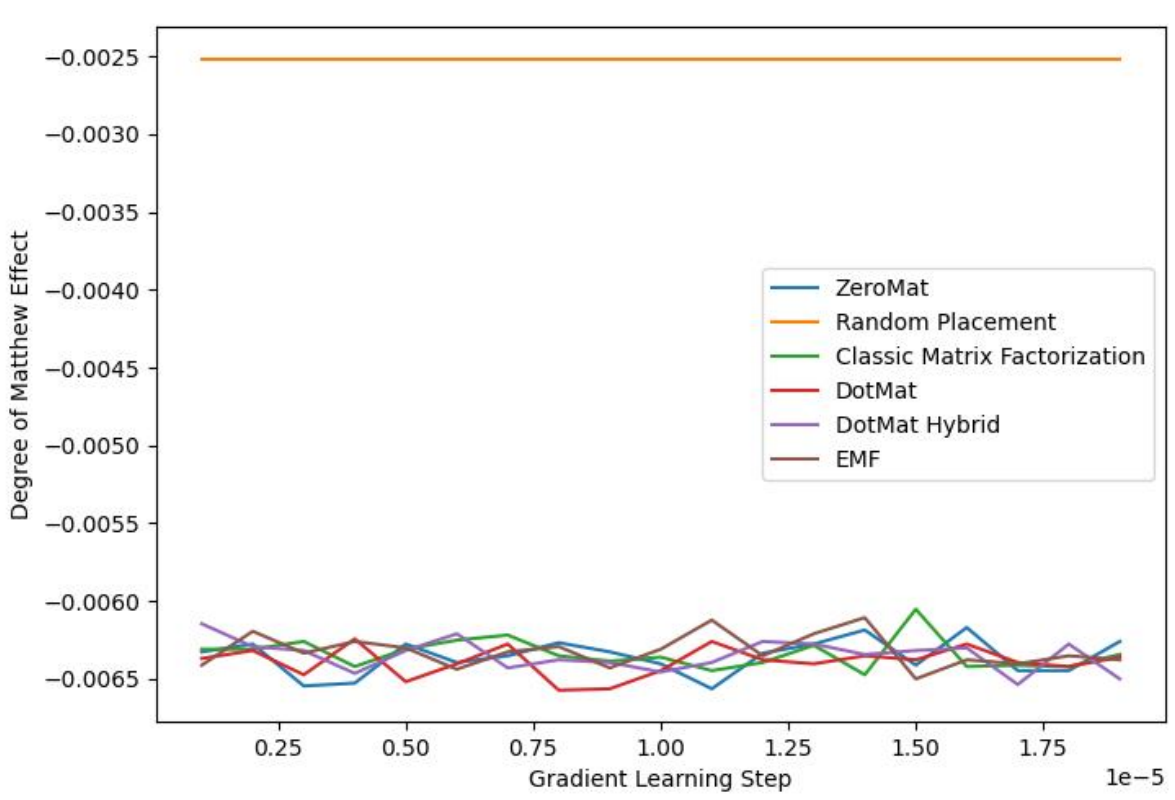

Figure 4 Comparison of algorithms on MovieLens 1 Million Dataset (Degree of Matthew Effect)

From Figure 3 and Figure 4, we observed that emotion-based recommender system does not necessarily lead to the best performance among recommendation algorithms on accuracy metric, but it does lead to better performance than others on the fairness metric.

From Figure 5, we observed that when testing on a different dataset (LDOS-CoMoDa), emotion-based approach is more competitive than before. Fairness analysis of this dataset is analogous to MovieLens Dataset, so we omit the details here.

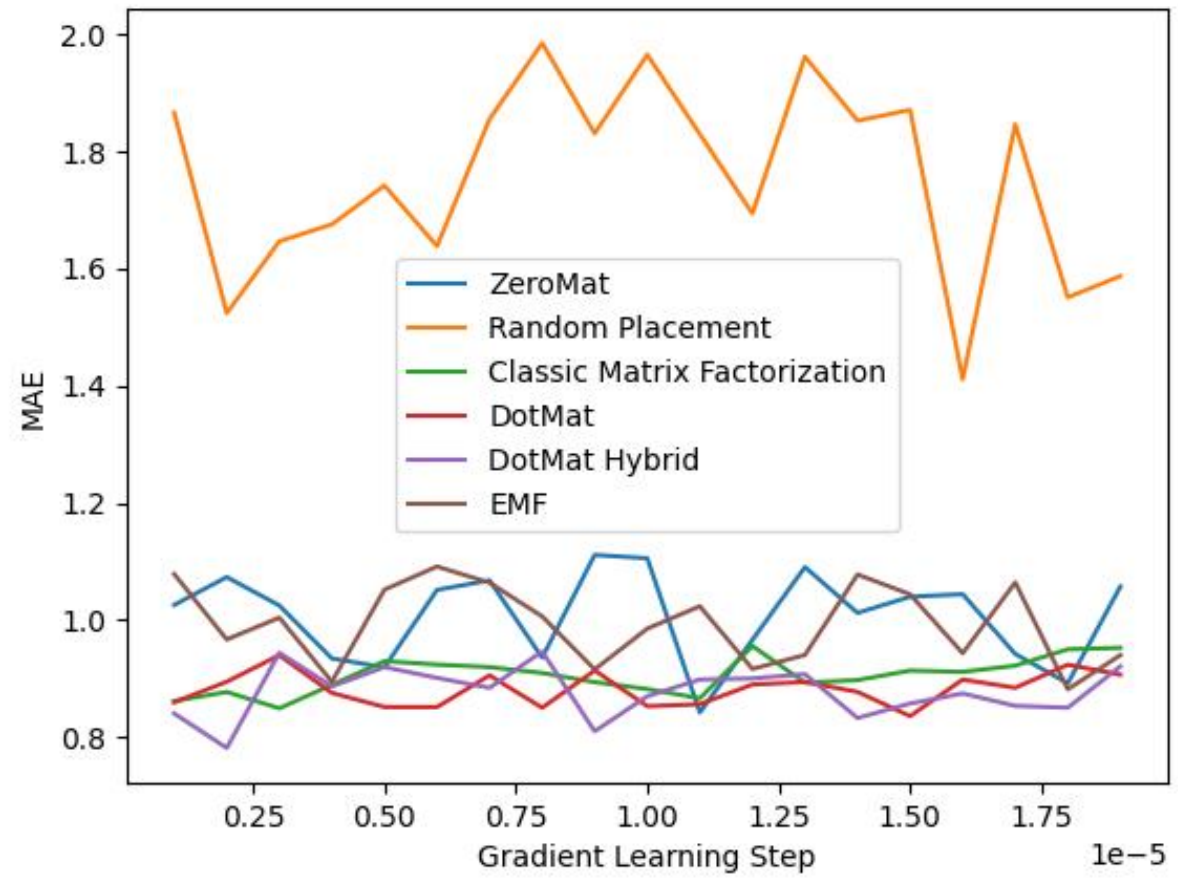

Figure 5 Comparison of algorithms on LDOS-CoMoDa Dataset (MAE)

## VII. CONCLUSION

In this paper we introduced a new theory and metric to represent users' emotion in recommender systems. We visualized data and analyzed the results on famous open datasets. In the end, we provide a framework to build emotion-aware recommender system and use emotion-based matrix factorization as an example to demonstrate the effectiveness of our theory.

In future work, we would like to explore how to increase the performance of accuracy using emotion-based recommender systems.

## ACKNOWLEDGMENT

After clinching 5 best paper award / best oral presentation awards from international research conferences (CCF-C and EI Compendex), I made a paper through at an SCI-indexed journal this year. This achievement is made 16 years after when I first published at an SCI-indexed journal (IEEE TVCG, CCF-A). The attainment might seem trivial for a researcher affiliated with top universities, but since I'm an independent researcher on my own, it means a great deal to me.

The idea of this paper struck on me sometime in 2024, but I wasn't able to think through it clearly so it can be materialized into a research paper until today. Despite the fact that this paper is based on simple mathematics, I believe it is going to be the first step towards emotional AI.